%% file: main.tex
\documentclass[11pt,letterpaper]{article}
\input{preamble}

\title{On the Limitations of Truthful\\Maximin Share Approximation}
\author{Moshe Babaioff\thanks{The Hebrew University of Jerusalem (HUJI). Email: \texttt{moshe.babaioff@mail.huji.ac.il}. Supported by a Golda Meir Fellowship and the Israel Science Foundation (grant No. 301/24).}
\qquad Gefen Frosh\thanks{The Hebrew University of Jerusalem (HUJI). Email: \texttt{gefen.frosh@mail.huji.ac.il}. Supported by the Israel Science Foundation (grant No. 301/24).}}
\date{November 27, 2026}

\begin{document}
\maketitle
\input{sections/abstract}
\input{sections/introduction}
\input{sections/model}
\input{sections/two-agents}
\input{sections/three-agents}
\input{sections/general-goods}
\input{sections/ai-disclosure}
\printbibliography
\end{document}

%% file: preamble.tex
\usepackage[margin=1in]{geometry}
\usepackage[backend=biber,style=alphabetic,maxnames=99,minnames=99,natbib=true]{biblatex}
\usepackage{amsmath,amssymb,amsthm}
\usepackage{booktabs}
\usepackage{enumitem}
\usepackage{etoolbox}
\usepackage{titlesec}
\titlespacing*{\paragraph}{0pt}{5pt}{1em}
\usepackage{microtype}
\usepackage[hyperfootnotes=false,colorlinks=true,linkcolor=blue,citecolor=blue,urlcolor=blue]{hyperref}
\hypersetup{pdftitle={On the Limitations of Truthful Maximin Share Approximation},pdfauthor={Moshe Babaioff and Gefen Frosh}}

\setlist[itemize]{itemsep=4pt,topsep=4pt}
\AtBeginEnvironment{tabular}{\small}
\AtBeginBibliography{\setlength{\emergencystretch}{2em}}
\newtheorem{theorem}{Theorem}
\newtheorem{lemma}{Lemma}
\newcommand{\E}{\mathbb{E}}
\makeatletter
\renewcommand{\maketitle}{%
  \begingroup
  \renewcommand{\thefootnote}{\fnsymbol{footnote}}
  \begin{center}
    {\LARGE\bfseries\@title\par}
    \ifx\@author\@empty\else\vspace{8pt}{\large\@author\par}\fi
    \vspace{10pt}
    {\small\@date\par}
  \end{center}
  \@thanks
  \endgroup
  \setcounter{footnote}{0}
}
\makeatother

%% file: sections/abstract.tex
\begin{abstract}
We study the allocation of indivisible goods among $n$ equally entitled, risk-neutral agents with nonnegative additive valuations, without payments. We work in a prior-free (non-Bayesian) model: no probability distribution over valuations is assumed. We require dominant-strategy truthfulness, namely truthfulness in expectation (TIE) for randomized mechanisms, together with a maximin share (MMS) guarantee in every realized allocation. Recent work gives a TIE mechanism with a constant MMS approximation guarantee for any number of agents: Ma, Tao, and Wang~\cite{ma-tao-wang} guarantee $\frac{1}{6}$ of the truncated proportional share, and hence of the MMS. Without incentive constraints, Feige and Grinberg~\cite{feige-grinberg} and Lin et al.~\cite{lin-et-al-mms} show that the best MMS approximation converges to one as the number of agents grows. Can TIE mechanisms achieve the same MMS guarantees as algorithms for nonstrategic agents?

We show that they cannot. For every $n\geq3$, no TIE direct-revelation mechanism guarantees more than $\frac{9}{10}$ of each agent's MMS, even with free disposal. In fact, this impossibility holds even when all agents share a publicly known strict ranking of the goods.
For unrestricted additive reports, we obtain the stronger upper bound $\frac{6}{7}$ for three agents. For two agents we prove an upper bound of $\frac{2}{3}$ of the MMS,
matching the TIE guarantee of Babaioff, Feige, and Manaker Morag~\cite{babaioff-et-al} and establishing the optimal approximation factor. Our general goods construction uses only $n+5$ goods. Exact MMS allocations always exist in this range by Feige, Sapir, and Tauber~\cite{feige-sapir-tauber}; thus the loss is due to incentives alone. For unrestricted reports, we also derive quantitative lower bounds on the incentive slack needed to exceed these approximation factors.
All our constructions are based on the chore constructions of Lin, Tao, Wu, and Zhang~\cite{lin-et-al-chores}, adapted here to goods.
\end{abstract}

%% file: sections/introduction.tex
\section{Introduction}\label{sec:introduction}
We consider the problem of allocating a set of indivisible goods among $n$ equally entitled, risk-neutral agents with nonnegative additive valuations. A basic goal is to give every agent a bundle whose value is at least her fair share. Indivisibility makes this goal difficult: even two agents who value a single good equally cannot both receive a positive fraction of its value in a deterministic allocation.

The maximin share benchmark addresses indivisibility through the principle of equal treatment of equals: it is the largest value that could be guaranteed to every agent if all agents had the same valuation.

\paragraph{The maximin share benchmark.}
The \emph{maximin share} (MMS), introduced by Budish~\cite{budish}, is the value an agent can secure by partitioning the goods into $n$ bundles and receiving a least-valued bundle. For two agents, cut-and-choose guarantees both agents their MMS. With three or more agents, however, exact MMS allocations need not exist~\cite{kurokawa-procaccia-wang}. This has led to an extensive study of approximate MMS allocations: an $\alpha$-MMS allocation gives every agent at least an $\alpha$ fraction of her MMS. The best known guarantee that holds uniformly for all numbers of agents is $\frac{7}{9}$~\cite{huang-zhou}.

A different picture emerges when the number of agents is large. Feige and Grinberg~\cite{feige-grinberg} and, independently, Lin, Liu, Tao, Wu, and Zhang~\cite{lin-et-al-mms} prove that for every fixed $\epsilon>0$, $(1-\epsilon)$-MMS allocations exist for all sufficiently large $n$. Thus, although exact MMS may be infeasible, the loss from approximating it vanishes as the number of agents grows.

\paragraph{Private valuations and incentives.}
These guarantees concern the values supplied to the allocation algorithm. When valuations are private, an agent may benefit by reporting different values, and an allocation that is fair with respect to the reports need not be fair with respect to the true valuations. We work in a \emph{prior-free} model, with no assumed distribution over valuations and no payments. We require \emph{truthfulness in expectation} (TIE): for every fixed profile of the other agents' reports, truthful reporting maximizes an agent's expected value, where the expectation is only over the mechanism's randomization. The MMS guarantee must hold at every valuation profile and \emph{ex post}, in every allocation in the mechanism's support. This combination of TIE and ex-post share guarantees has been studied by Bu and Tao~\cite{bu-tao}, Babaioff, Feige, and Manaker Morag~\cite{babaioff-et-al}, and Ma, Tao, and Wang~\cite{ma-tao-wang}.

These works establish positive TIE approximation guarantees; we focus on upper bounds for unrestricted TIE mechanisms. Prior impossibilities under additional information restrictions are discussed below.

Deterministic truthfulness imposes strong restrictions. For such mechanisms, Amanatidis, Birmpas, and Markakis~\cite{amanatidis-2016} establish impossibilities for MMS approximation in several information models. For two agents, Amanatidis et al.~\cite{amanatidis-2017} characterize the case in which every good must be allocated (no free disposal) and show that the best deterministic MMS approximation factor decreases to zero as the number of goods grows large.

Randomization permits substantially stronger guarantees. Bu and Tao~\cite{bu-tao} obtain $\frac{1}{n}$-MMS under TIE. Babaioff, Feige, and Manaker Morag~\cite{babaioff-et-al} improve the guarantee to $\Theta(\frac{1}{\log n})$ using ordinal information and obtain $\frac{2}{3}$-MMS for two agents. Ma, Tao, and Wang~\cite{ma-tao-wang} subsequently obtain a constant $\frac{1}{6}$ guarantee for arbitrary $n$. The last two works prove their guarantees for the \emph{truncated proportional share} (TPS), introduced by Babaioff, Ezra, and Feige~\cite{babaioff-ezra-feige-bobw}, which is at least the MMS.

\paragraph{The remaining gap.}
The constant MMS approximation guarantee of TIE mechanisms and the asymptotically almost-perfect approximation of nonstrategic algorithms raise a natural question:
\begin{quote}
\emph{Can dominant-strategy truthful randomized mechanisms (TIE mechanisms) achieve the same MMS approximation as arbitrary allocation algorithms?}

\end{quote}
The known deterministic and ordinal impossibilities do not answer this question. A truthful mechanism may use both full valuation reports and randomization. To bound what it can achieve, one must distinguish the restrictions imposed by incentives from the possible nonexistence of a fair allocation.

We give a negative answer, with the bounds described below.

\subsection{Our Results}\label{sec:our-results}
We present results for TIE mechanisms when allocating goods to $n$ additive agents, when there are no transfers. 
All our bounds allow free disposal and require no anonymity, neutrality, non-bossiness, efficiency, or computational assumption.

\paragraph{A constant upper bound for arbitrary numbers of agents.}
For every $n\geq3$, we show that no TIE mechanism guarantees more than $\frac{9}{10}$-MMS ex post. Together with the asymptotic existence results~\cite{feige-grinberg,lin-et-al-mms}, this establishes a constant separation between the best guarantees with and without incentive constraints, even as $n$ tends to infinity.

The construction uses only $n+5$ goods. Feige, Sapir, and Tauber~\cite{feige-sapir-tauber} prove that exact MMS allocations exist for every additive profile with at most $n+5$ goods. Consequently, on this item domain the entire loss is caused by incentives, for each fixed $n\geq3$. For unrestricted item sets, comparison with the known nonstrategic guarantees establishes separation for $n=2,3$ and all sufficiently large $n$, as summarized in Section~\ref{sec:model}.

The $\frac{9}{10}$ bound holds even when all agents share a publicly known strict ranking and every report must respect it. We consider the special case of the public-rankings model of Amanatidis, Birmpas, and Markakis~\cite[Section 5]{amanatidis-2016} in which all agents share the same strict ranking. Only the cardinal values are private.

\paragraph{Stronger bounds for few agents with unrestricted reports.}
For two agents, we prove a $\frac{2}{3}$ upper bound on the MMS approximation of TIE mechanisms, using four goods and three reports. This matches the TIE mechanism of Babaioff, Feige, and Manaker Morag~\cite{babaioff-et-al}, so the optimal factor is exactly $\frac{2}{3}$. Their mechanism also guarantees $\frac{2}{3}$-TPS, which is optimal for TPS even without incentive constraints. Their Proposition 4 also bounds MMS approximation by $\frac{2}{3}$ when the induced fractional allocation rule is ordinal, even without truthfulness~\cite{babaioff-et-al}. Our impossibility establishes the MMS upper bound for TIE mechanisms without that ordinal restriction.
For three agents, we obtain the stronger upper bound $\frac{6}{7}$ by adapting the three-report incentive chain used for chores by Lin, Tao, Wu, and Zhang~\cite[Section 5.2]{lin-et-al-chores}.

\paragraph{Approximate truthfulness.}
For unrestricted additive reports, we quantify how much multiplicative incentive slack is necessary to exceed each of the $\frac{2}{3}$, $\frac{6}{7}$, and $\frac{9}{10}$ bounds. In particular, slack $o(\frac{1}{n})$ cannot support a guarantee exceeding $\frac{9}{10}$ by a fixed positive constant as $n$ grows.

\subsection{Proof Approach}\label{sec:techniques}
All our constructions are based on those of Lin, Tao, Wu, and Zhang~\cite[Sections 5.1--5.3]{lin-et-al-chores} for chores. We adapt their three-report incentive chains for two and three agents and their single-deviation argument for general $n$ to goods. Ex-post fairness restricts the possible bundles, and incentive constraints compare their expected values across reports. Tracking incentive slack gives the approximate-TIE bounds; a perturbation and compactness argument gives the public-ranking strengthening.

\input{sections/related-work}

\subsection{Organization}
Section~\ref{sec:model} defines the model and summarizes the bounds. Section~\ref{sec:two-agents} proves the tight two-agent bound, Section~\ref{sec:three-agents} proves the three-agent bound, and Section~\ref{sec:general-goods} proves the general bound and its public-ranking strengthening.

%% file: sections/related-work.tex
\subsection{Additional Related Work}\label{sec:related-work}
We discuss the lines of work most closely connected to our bounds, complementing the TIE results described above.

\paragraph{MMS approximation without incentives.}
Following the introduction of MMS by Budish~\cite{budish}, Procaccia and Wang~\cite{procaccia-wang} established a $\frac{2}{3}$ guarantee for additive goods. Subsequent work developed approximation algorithms and improved the guarantees~\cite{amanatidis-approximation,barman-krishnamurthy,ghodsi-et-al,garg-mcglaughlin-taki,garg-taki,akrami-garg-sharma-taki,akrami-garg,heidari-et-al}, culminating in the uniform $\frac{7}{9}$ bound of Huang and Zhou~\cite{huang-zhou}. For few agents, Feige and Norkin~\cite{feige-norkin} obtain $\frac{11}{12}$ for three agents, and Ghodsi et al.~\cite[Theorem A.6]{ghodsi-et-al} obtain $\frac{4}{5}$ for four. Kurokawa, Procaccia, and Wang~\cite{kurokawa-procaccia-wang} show that exact MMS need not exist; Feige, Sapir, and Tauber~\cite{feige-sapir-tauber} strengthen the negative examples and establish the $n+5$-goods existence result used here. For three agents, their nine-good example rules out guarantees above $\frac{39}{40}$-MMS even without incentives~\cite{feige-sapir-tauber}. 
For general $n$, Feige, Sapir, and Tauber~\cite{feige-sapir-tauber} establish an upper bound of $1-\frac{1}{n^4}$ for $n\geq4$. Ezra and Garbuz~\cite[Theorem 1]{ezra-garbuz} strengthen this to $1-\Omega(\frac{1}{(\log n)^2})$ for all sufficiently large $n$, even without incentive constraints.
For TIE mechanisms, we prove the asymptotically stronger upper bound of $\frac{9}{10}$ for every $n\geq3$: the loss remains at least $\frac{1}{10}$ as $n$ grows. Without incentives, the optimal approximation converges to one~\cite{feige-grinberg,lin-et-al-mms}, giving a strict asymptotic separation.

\paragraph{Information and truthfulness.}
Amanatidis, Birmpas, and Markakis~\cite{amanatidis-2016} distinguish cardinal, ordinal, and public-ranking models. Their upper bound for deterministic ordinal algorithms is $\frac{1}{H_n}$, even without incentives, where $H_n=\sum_{j=1}^n\frac{1}{j}$ is the $n$-th harmonic number; Halpern and Shah~\cite{halpern-shah} study the fairness guarantees obtainable from partial rankings. For two agents and $m$ goods, the optimal deterministic guarantee under complete allocation is $\frac{1}{\left\lfloor\frac{\max\{2,m\}}{2}\right\rfloor}$~\cite[Application 4.7]{amanatidis-2017}. Babaioff and Manaker Morag~\cite{babaioff-manaker-morag} study truthful deterministic mechanisms for arbitrary numbers of agents under non-bossiness and neutrality.
Their characterization allows free disposal. For randomized mechanisms, Babaioff, Feige, and Manaker Morag~\cite{babaioff-et-al} give a $\Theta(\frac{1}{\log n})$ ex-post MMS guarantee under TIE, and Ma, Tao, and Wang~\cite{ma-tao-wang} give a constant $\frac{1}{6}$ guarantee.
Our TIE bounds for unrestricted additive valuations impose neither non-bossiness nor neutrality and do not require complete allocation of goods.

\paragraph{Restricted domains and randomized mechanisms.}
Truthfulness can coexist with strong fairness guarantees on restricted domains, including binary additive and matroid-rank valuations~\cite{halpern-et-al,babaioff-ezra-feige-dichotomous}. For divisible resources, Mossel and Tamuz~\cite{mossel-tamuz} and Chen et al.~\cite{chen-et-al} study truthful fair division, while Freeman et al.~\cite{freeman-et-al} relate truthful fractional allocation rules to wagering mechanisms. These results motivate the use of randomization, but do not provide ex-post MMS guarantees for arbitrary indivisible-goods instances. Garg, Narayan, and Shen~\cite{garg-narayan-shen} give a TIE mechanism whose allocation is envy-free with high probability under random-value assumptions as the number of goods grows. Their incentive guarantee is TIE for every fixed profile of opposing reports, with expectation only over the mechanism's randomness~\cite[Section 2.1.2]{garg-narayan-shen}; the distributional assumptions support the high-probability fairness guarantee, not merely Bayesian incentive compatibility. Our asymptotic question is different: the number of agents grows, and the guarantee must hold at every additive profile and in every supported allocation.

\paragraph{Best-of-Both-Worlds fairness.}
Best-of-Both-Worlds (BoBW) algorithms combine fairness in expectation with fairness in every realized allocation. Aziz et al.~\cite{aziz-et-al-bobw} combine ex-ante envy-freeness with ex-post envy-freeness up to one good (EF1). Babaioff, Ezra, and Feige~\cite{babaioff-ezra-feige-bobw} introduce TPS and combine ex-ante proportionality with ex-post $\frac{1}{2}$-TPS. Babaioff and Frosh~\cite{babaioff-frosh} obtain stronger guarantees for few agents, including ex-ante proportionality and ex-post $\frac{9}{10}$-MMS for three agents. Such distributional fairness conditions do not imply TIE. In particular, our $\frac{6}{7}$ ceiling for three agents separates the achievable BoBW guarantee from the TIE guarantee even without imposing any ex-ante fairness requirement on the latter.

\paragraph{Approximate incentives.}
Xiao and Ling~\cite{xiao-ling} and Tao and Yang~\cite{tao-yang} study multiplicative incentive ratios in fair allocation. Babaioff, Feige, and Manaker Morag~\cite{babaioff-et-al} obtain an $\Omega(\frac{1}{\log\log n})$ ex-post guarantee with limited cardinal information and multiplicative TIE slack $n^{-\log n}$. Our bounds concern the slack necessary to reach a substantially higher MMS factor; they do not contradict that almost-TIE guarantee. Share guarantees also give a weaker form of incentive protection: reporting truthfully secures the promised share independently of the other reports. Babaioff and Feige~\cite{babaioff-feige-shares} study this distinction through self-maximizing shares. Securing a share does not in itself make truth-telling maximize expected utility.

\paragraph{Construction credit.}
All our constructions are based on the chore constructions of Lin, Tao, Wu, and Zhang~\cite[Sections 5.1--5.3]{lin-et-al-chores}. The two-agent proof adapts their four-item, three-report construction from Section 5.1, using different goods valuations and incentive comparisons. The three-agent proof adapts the construction and incentive chain of their Section 5.2. Our general goods proof uses the homogeneous-profile and single-deviation method of their Section 5.3, with an eight-good instance and the corresponding bundle argument. Their chore instances and exact-TIE lower bounds are due to that work.

%% file: sections/model.tex
\section{Model and statement of results}\label{sec:model}
We consider a finite set $G$ of $m=|G|$ indivisible goods and $n$ equally entitled agents. Agent $i$ has a nonnegative additive valuation $v_i$: the value of a bundle is the sum of its item values. Agents are risk-neutral and maximize expected value. We work in a prior-free (non-Bayesian) model: valuations are arbitrary, and no prior distribution over them is assumed. All expectations are over the mechanism's internal randomization, not over agents' valuations. There are no payments. Unless explicitly restricted, the report domain consists of all nonnegative additive valuations.

\paragraph{Allocations and mechanisms.}
An allocation consists of pairwise disjoint bundles $(X_1,\ldots,X_n)$, one for each agent. Goods may be discarded, so their union need not equal $G$. A mechanism maps each profile of additive reports to a lottery over allocations. Every MMS guarantee required of the mechanisms in our impossibility results is \emph{ex post}: it must hold in every allocation in the support of that lottery.

\paragraph{Truthfulness in expectation.}
Write $X_i(v_i,v_{-i})$ for agent $i$'s random bundle when she reports $v_i$ and the other agents report $v_{-i}$. A goods mechanism is \emph{truthful in expectation} (TIE) if
\[
\E[v_i(X_i(v_i,v_{-i}))]\geq\E[v_i(X_i(v'_i,v_{-i}))]
\]
for every agent $i$, true valuation $v_i$, alternative report $v'_i$,
and fixed profile $v_{-i}$ of the other agents' reports. In particular, this inequality must hold separately at each such profile, rather than only after averaging over a distribution of $v_{-i}$. We impose no anonymity, neutrality, non-bossiness, or computational assumption.

\paragraph{Scope of truthfulness.}
Throughout, \emph{truthful} means dominant-strategy truthful for risk-neutral agents, hence TIE for randomized mechanisms. Deterministic truthful mechanisms and universally truthful mechanisms (report-independent lotteries over deterministic truthful mechanisms) are special cases~\cite{babaioff-et-al}. Thus, our TIE impossibilities also apply to these stronger notions. Agents may report a numerical value for every good, rather than only a ranking. TIE must hold for every admissible change to these reported values.
An indirect mechanism with dominant strategies can be represented directly by mapping each reported type to its prescribed dominant strategy, preserving its allocation lotteries. Our bounds therefore also rule out such implementations. Bayesian incentive compatibility instead averages over other agents' types under a specified prior. It is a weaker requirement, and our impossibility results do not rule out mechanisms satisfying only that requirement.

\paragraph{Maximin shares.}

For an item set $G$ and $n$ agents, the MMS benchmark is
\[
\mu_v=\max_{(P_1, P_2, \ldots,P_n)}\min_j v(P_j),
\]
over partitions $(P_1,P_2,\ldots,P_n)$ of $G$ into $n$ bundles, allowing empty bundles. An $\alpha$-MMS allocation gives each agent $i$ value at least $\alpha\cdot \mu_{v_i}$.

\begin{theorem}\label{thm:goods}
Consider settings with $n$ equally entitled, risk-neutral agents with nonnegative additive valuations over indivisible goods, and mechanisms without payments.
The optimal ex-post MMS approximation factor of a TIE mechanism, even allowing free disposal, is $\frac{2}{3}$ for two agents, at most $\frac{6}{7}$ for three agents, and at most $\frac{9}{10}$ for every $n\geq4$.
The $\frac{9}{10}$ upper bound holds for every $n\geq3$ even when all agents share a publicly known strict ranking and all reports must respect that ranking.
\end{theorem}
\begin{center}
\renewcommand{\arraystretch}{1.4}
\begin{tabular}{c|cc|l}
\toprule
$n$&Nonstrategic guarantee&TIE upper bound&Separation implied?\\\midrule
$2$&$1$&$\frac{2}{3}$&Yes\\
$3$&$\frac{11}{12}$~\cite{feige-norkin}&$\frac{6}{7}$&Yes\\
$4$&$\frac{4}{5}$~\cite{ghodsi-et-al}&$\frac{9}{10}$&--\\
\shortstack{$n\geq5$\\(uniform guarantee)}&$\frac{7}{9}$~\cite{huang-zhou}&$\frac{9}{10}$&--\\
\shortstack{Sufficiently large $n$\\(asymptotic guarantee)}&$\frac{19}{20}$~\cite{feige-grinberg,lin-et-al-mms}&$\frac{9}{10}$&Yes\\
\bottomrule
\end{tabular}
\end{center}
The table compares known MMS guarantees without incentives with our upper bounds for TIE mechanisms. These bounds give a strict separation for two agents, three agents, and sufficiently many agents; for the remaining cases, marked by a dash, they do not establish a separation. The last row follows from the asymptotic guarantees of~\cite{feige-grinberg,lin-et-al-mms}. The table allows unrestricted additive valuations, but our $\frac{9}{10}$ upper bound for $n\geq3$ holds even with a publicly known common strict ranking. Moreover, for every $n\geq2$, we obtain a separation on a fixed set of goods for which exact MMS allocations always exist (Section~\ref{sec:general-goods}).

The factors above are worst-case guarantees over all finite item sets and all profiles in the stated domain, rather than claims for every fixed number of items. Our upper bounds use four goods for two agents, nine for the $\frac{6}{7}$ bound, and $n+5$ for the $\frac{9}{10}$ bound. For unrestricted reports, additional zero-valued goods preserve these impossibilities. For the common-strict-ranking result, append the zero-valued goods before applying the perturbation in Section~\ref{sec:general-goods} (the paragraph on public rankings).

\paragraph{Quantitative truthfulness.}
The proofs also bound multiplicative incentive slack $0\leq\delta<1$: truthful expected utility must be at least $(1-\delta)$ times the expected utility from any misreport, for each fixed profile of the other agents' reports. Exact TIE corresponds to $\delta=0$. For the additive-goods model above, with free disposal, the following inequalities are necessary for the indicated ex-post approximation factors:
\begin{center}
\renewcommand{\arraystretch}{1.8}
\begin{tabular}{c|c|c}
Setting&Approximation factor&Necessary incentive slack\\\hline
Goods, $n=2$&$\frac{2}{3}<\alpha\leq1$&$\displaystyle\delta\geq\frac{3\alpha-2}{18+6\alpha}$\\
Goods, $n=3$&$\frac{6}{7}<\alpha\leq1$&$\displaystyle\delta\geq\frac{2\eta}{102+\eta},\quad\eta=7-\frac{6}{\alpha}$\\
Goods, $n\geq3$&$\frac{9}{10}<\alpha\leq1$&$\displaystyle\delta\geq\frac{10\alpha-9}{10\alpha n}$
\end{tabular}
\end{center}
In particular, slack $o(\frac{1}{n})$ cannot remove the asymptotic constant gap.

%% file: sections/two-agents.tex
\section{Two agents: \texorpdfstring{$\frac{2}{3}$}{2/3} MMS is best possible for TIE mechanisms}\label{sec:two-agents}
For two agents, ordinary cut-and-choose achieves exact MMS. For additive goods, we show that truthfulness reduces the best factor to $\frac{2}{3}$.

Babaioff, Feige, and Manaker Morag~\cite[Theorem 3]{babaioff-et-al} give the matching positive result: their mechanism guarantees $\frac{2}{3}$-TPS, and hence $\frac{2}{3}$-MMS. They also prove a $\frac{2}{3}$-MMS upper bound for mechanisms whose induced fractional allocation rule is ordinal, even without truthfulness~\cite[Proposition 4]{babaioff-et-al}. Our upper bound removes that information restriction for TIE mechanisms, allowing arbitrary cardinal reports. We allow free disposal throughout. The proof first bounds the value obtainable by an auxiliary report against a fixed opponent, and then uses the homogeneous profile to obtain a contradiction.

Our construction is based on the four-item, three-report construction of Lin, Tao, Wu, and Zhang~\cite[Section 5.1]{lin-et-al-chores} for chores. We use different goods valuations and incentive comparisons to obtain the $\frac{2}{3}$-MMS upper bound.

Fix $\frac{2}{3}<\alpha\leq1$, and choose $0<\varepsilon<\min(\alpha-\frac{2}{3},\frac{1}{3})$. The three reports on four goods are:
\begin{center}
\renewcommand{\arraystretch}{1.3}
\begin{tabular}{lrrrrr}
\toprule
&Good 1&Good 2&Good 3&Good 4&MMS\\\midrule
$A$&$1$&$3$&$1$&$1$&$3$\\
$C$&$3$&$1$&$1$&$1$&$3$\\
$B$&$\frac{2}{3}+\varepsilon$&$\frac{2}{3}$&$\frac{1}{3}$&$\frac{1}{3}$&$1$\\\bottomrule
\end{tabular}
\end{center}
For $A$ and $C$, separating the value-$3$ good from the other three goods gives MMS $3$, which also equals half the total value. For $B$, the partition $\{1,3\}\mid\{2,4\}$ gives minimum value $1$. No partition gives both bundles value above $1$: a singleton is worth less than $1$, and in a split into pairs the pair without good $1$ is worth at most $1$. Thus $\mu_B=1$.

Since $3\alpha>2$ and $A,C$ have integer item values, a fair bundle for either report has value at least $3$. Every singleton has $B$-value below $\alpha$, and $B(\{3,4\})=\frac{2}{3}<\alpha$. Hence a fair $B$-bundle contains at least two goods, including good $1$ or $2$.

At $(A,C)$ both agents must get their value-$3$ good. Indeed, an agent missing her value-$3$ good would need all three other goods, leaving her opponent only the good she values at $1$. At $(B,C)$, $C$ must get good $1$ (else $B$ cannot get two goods), so $B$ gets good $2$ and at least one of goods $3$ and $4$. The analogous claim holds at $(A,B)$.

\paragraph{Roadmap of the incentive comparisons.}
We first prove the impossibility for exact TIE. In the table, each arrow changes only one agent's report; both outcomes are evaluated using that agent's \emph{true valuation}, listed in the second column. Profiles are ordered by agent identity. Here $u_A,u_C$ are the respective truthful expected utilities at $(A,C)$, and $k$ counts the goods from $\{3,4\}$ received by the $B$-reporter at $(B,C)$.
\begin{center}
\renewcommand{\arraystretch}{1.3}
\begin{tabular}{p{0.06\linewidth}p{0.16\linewidth}p{0.30\linewidth}p{0.36\linewidth}}
\toprule
Step&True valuation&Comparison or calculation&Bound obtained\\\midrule
1&$A$ (agent 1)&$(A,C)\to(B,C)$&TIE gives $u_A\geq4$.\\
1&$C$ (agent 2)&$(A,C)\to(A,B)$&TIE gives $u_C\geq4$.\\
1&$A$ and $C$&Total utility at $(A,C)$ is at most $8$.&Together, $u_A=u_C=4$.\\\midrule
2&$A$ (agent 1)&$(A,C)\to(B,C)$&$3+\E[k]\leq4$, so $\E[k]\leq1$.\\
2&$B$ (either agent)&Value the truthful bundle against $C$ using $B$.&Truthful utility is $\frac{2}{3}+\frac{\E[k]}{3}\leq1$.\\\midrule
3&$B$ (selected agent)&Report $C$ instead of $B$, facing $C$.&Deviation utility is at least $1+\frac{\varepsilon}{2}$.\\
3&Same true $B$&TIE requires truthful utility $\geq$ deviation utility.&Would require $1\geq1+\frac{\varepsilon}{2}$: a contradiction.\\
\bottomrule
\end{tabular}
\end{center}
Step 2 uses the incentive constraint of a true $A$-agent to bound the truthful utility of a $B$-agent facing $C$. Step 3 shows that some such $B$-agent gains by reporting $C$.

\paragraph{Step 1: both truthful utilities at $(A,C)$ equal $4$.}
Let $u_A$ and $u_C$ be the agents' expected utilities at $(A,C)$, evaluated using their respective true valuations. A true $A$-agent can report $B$ and receive good $2$ and at least one of goods $3,4$, worth at least $4$ under $A$. Likewise, a true $C$-agent can report $B$ and receive good $1$ and at least one of goods $3,4$, worth at least $4$ under $C$. TIE implies $u_A,u_C\geq4$.

At $(A,C)$, each agent receives her value-$3$ good, and goods $3,4$ contribute at most $2$ in total. Thus $u_A+u_C\leq8$, even with disposal. Consequently,
\[
 u_A=u_C=4.
\]

\paragraph{Step 2: a truthful $B$-agent facing $C$ receives value at most $1$.}
Let $Y$ be agent $1$'s random bundle at $(B,C)$. It contains good $2$, excludes good $1$, and contains a random number $k\in\{1,2\}$ of goods from $\{3,4\}$. The key is to distinguish its two valuations:
\[
 A(Y)=3+k,\qquad B(Y)=\frac{2}{3}+\frac{k}{3}.
\]
A \emph{true $A$-agent} facing $C$ receives $Y$ by reporting $B$. Her incentive constraint, together with Step 1, gives
\[
 3+\E[k]=\E[A(Y)]\leq u_A=4,
 \qquad\text{so}\qquad \E[k]\leq1.
\]
Now evaluate the same lottery using $B$. A \emph{true $B$-agent} facing $C$ obtains it by reporting truthfully, so her expected utility is
\[
 \E[B(Y)]=\frac{2}{3}+\frac{\E[k]}{3}\leq1.
\]
Repeating Steps 1 and 2 with the agent identities exchanged, starting from $(C,A)$, gives this bound for either named agent facing $C$. No anonymity is assumed.

\paragraph{Step 3: some true $B$-agent gains by reporting $C$.}
At $(C,C)$, the agent without good $1$ needs all three remaining goods to reach value $3$. Every fair allocation therefore splits the goods as $\{1\}\mid\{2,3,4\}$. Choose a named agent whose probability $p$ of receiving $\{1\}$ is at most $\frac{1}{2}$.

Suppose this agent's true valuation is $B$ and her opponent reports $C$. By reporting $C$, she induces $(C,C)$. Since
\[
 B(\{1\})=\frac{2}{3}+\varepsilon<\frac{4}{3}
 =B(\{2,3,4\}),
\]
her expected true value from this deviation is at least
\[
 p\left(\frac{2}{3}+\varepsilon\right)+(1-p)\frac{4}{3}
 \geq\frac{1}{2}\left(\frac{2}{3}+\varepsilon\right)
       +\frac{1}{2}\cdot\frac{4}{3}
 =1+\frac{\varepsilon}{2}.
\]
If this selected agent reports truthfully, she reports $B$ while her opponent still reports $C$. This is precisely the situation covered by Step 2, which applies to either named agent. Her bundle then consists of good $2$ and $k$ goods from $\{3,4\}$, with $\E[k]\leq1$. Her expected true $B$-value is therefore $\frac{2}{3}+\frac{\E[k]}{3}\leq1$. In contrast, reporting $C$ gives her the lottery at $(C,C)$ just evaluated. Hence
\[
 \underbrace{\E[B(\text{truthful bundle})]}_{\leq1}
 <\underbrace{\E[B(\text{bundle after reporting }C)]}_{\geq1+\frac{\varepsilon}{2}},
\]
contradicting TIE. This proves the $\frac{2}{3}$ upper bound.

\paragraph{Extension to approximate TIE.}
We now repeat the same comparisons with multiplicative slack $0\leq\delta<1$. The bundle restrictions above depend only on fairness and remain unchanged. Step 1 gives
\[
 u_A,u_C\geq4(1-\delta),\qquad u_A+u_C\leq8,
 \qquad\text{hence}\qquad u_A,u_C\leq4+4\delta.
\]
In Step 2, the true $A$-agent's constraint becomes
\[
 (1-\delta)(3+\E[k])\leq u_A\leq4+4\delta.
\]
Thus $\E[k]\leq1+\frac{8\delta}{1-\delta}$, and either named agent's truthful $B$-utility against $C$ is at most $1+\frac{8\delta}{3(1-\delta)}$. Step 3 still gives deviation value at least $1+\frac{\varepsilon}{2}$. The true $B$-agent's approximate-TIE constraint therefore requires
\[
 1+\frac{8\delta}{3(1-\delta)}
 \geq(1-\delta)\left(1+\frac{\varepsilon}{2}\right).
\]
Multiplying by $1-\delta>0$ and collecting terms yields
\[
 \frac{\varepsilon}{2}
 \leq\left(\frac{11}{3}+\varepsilon\right)\delta
       -\left(1+\frac{\varepsilon}{2}\right)\delta^2
 \leq\left(\frac{11}{3}+\varepsilon\right)\delta.
\]
Thus $\delta\geq\frac{3\varepsilon}{22+6\varepsilon}$. Letting $\varepsilon$ approach $\alpha-\frac{2}{3}$ from below gives
\[
 \delta\geq\frac{3\alpha-2}{18+6\alpha}.
\]
Free disposal was permitted throughout.\hfill$\square$

%% file: sections/three-agents.tex
\section{Three agents: an upper bound of \texorpdfstring{$\frac{6}{7}$}{6/7} MMS for  TIE mechanisms}\label{sec:three-agents}
The proof adapts the three-report construction and incentive chain of Lin et al.~\cite[Section 5.2]{lin-et-al-chores}, specialized to three agents. We retain their arrangement of one distinguished item, two large items, and six unit items, and choose goods valuations so that the ex-post guarantee forces minimum bundle values. The three steps below follow their sequence of deviations. We first prove the exact-TIE impossibility and then repeat the comparisons with incentive slack.

Fix $\alpha>\frac{6}{7}$, $\alpha\leq1$, and choose $\varepsilon\in(0,1)$ so that $\alpha(7-\varepsilon)>6$. The nine goods consist of one good $e$, two goods $b_1,b_2$, and six unit goods. Consider these three reports:
\begin{center}
\begin{tabular}{lrrrr|rr}
\toprule
Report&$e$&Each $b_j$&Unit good&MMS lower bound&\shortstack{Minimum unit goods\\needed alongside $e$}&\shortstack{Minimum unit goods\\needed alongside $b_j$}\\\midrule
$H$&$1$&$7$&$1$&$7$&$6$&$0$\\
$A$&$3-\varepsilon$&$6$&$1$&$7-\varepsilon$&$4$&$1$\\
$L$&$7$&$4$&$1$&$7$&$0$&$3$\\\bottomrule
\end{tabular}
\end{center}
The last two columns count the unit-valued goods an agent must receive in addition to her single special good ($e$ or $b_j$) for her bundle to have value strictly above $6$. For example, under $H$, good $e$ is worth $1$, so it must be accompanied by all six unit goods.
The indicated MMS lower bounds come from the following partitions: for $H$, put all six units with $e$ and leave each $b_j$ alone; for $A$, put four units with $e$ and one with each $b_j$; for $L$, leave $e$ alone and put three units with each $b_j$. Only these lower bounds, rather than exact MMS values, are needed below. Every fair bundle must have reported value above $6$, whereas all six unit goods together have value only $6$. Thus each agent must receive at least one of the three special goods $e,b_1,b_2$. Since there are three agents, each receives exactly one special good and at least the listed number of units.

\paragraph{Roadmap of the incentive comparisons.}
We first assume exact TIE. The table shows how the bounds lead to a profitable deviation. Each arrow changes one agent's report, and both outcomes are evaluated using her true valuation. Write $X_i$ for the bundles at $(L,H,H)$, $Y_i$ for the bundles at $(A,H,H)$, and $Z_i$ for the bundles at $(H,H,H)$.
\begin{center}
\renewcommand{\arraystretch}{1.3}
\begin{tabular}{p{0.06\linewidth}p{0.16\linewidth}p{0.30\linewidth}p{0.36\linewidth}}
\toprule
Step&True valuation&Comparison or calculation&Bound obtained\\\midrule
1&$H$ (agent 2)&$(L,H,H)\to(L,A,H)$&$\E[H(X_2)]\geq8$.\\
1&$H$ (agent 3)&$(L,H,H)\to(L,H,A)$&$\E[H(X_3)]\geq8$.\\
1&$L$ (agent 1)&$H(G)=21$ and $L(X_1)\leq H(X_1)+6$.&$\E[H(X_1)]\leq5$, hence $\E[L(X_1)]\leq11$.\\\midrule
2&$L$ (agent 1)&$(L,H,H)\to(A,H,H)$&$\E[L(Y_1)]\leq11$.\\
2&$A$ (agent 1)&Same lottery at $(A,H,H)$; $A(Y_1)=L(Y_1)-4-\varepsilon$.&Truthful utility $\E[A(Y_1)]\leq7-\varepsilon$. The same bound holds when any other agent reports $A$ against two $H$-reports.\\\midrule
3&$A$ (selected agent)&Report $H$ instead of $A$, facing $H,H$.&Some agent has deviation utility $\E[A(Z_i)]\geq7-\frac{\varepsilon}{3}$.\\
3&Same true $A$&TIE requires truthful utility $\geq$ deviation utility.&Would require $7-\varepsilon\geq7-\frac{\varepsilon}{3}$: a contradiction.\\
\bottomrule
\end{tabular}
\end{center}
Step 2 uses the same lottery at $(A,H,H)$ first to evaluate a deviation by a true $L$-agent and then to bound the truthful utility of a true $A$-agent. We explain the two comparisons below.

\paragraph{Step 1: bound a truthful $L$-agent's utility against two $H$-reports.}
First consider the true valuation profile $(L,H,H)$, and let $X_1,X_2,X_3$ be the random bundles when all three agents report truthfully. We begin with the incentive constraints of agents $2$ and $3$, whose true valuation is $H$.

If either of these agents reports $A$ instead, the reports become $(L,A,H)$ or $(L,H,A)$. In either case, the $A$-reporter cannot receive $e$: she would need four units, while the $L$-reporter would then receive a $b_j$ and need three more units. Only six units are available. Thus the deviating agent receives a $b_j$ and at least one unit. Evaluated using her \emph{true valuation $H$}, this bundle is worth at least $7+1=8$. TIE therefore implies
\[
 \E[H(X_2)]\geq8,
 \qquad \E[H(X_3)]\geq8.
\]
To bound what remains for agent $1$, evaluate \emph{all three bundles using $H$}. They are disjoint, and $H(G)=21$, so
\[
 \E[H(X_1)]
 \leq21-\E[H(X_2)]-\E[H(X_3)]
 \leq21-16=5.
\]
This is not yet agent $1$'s true utility: her valuation is $L$. However, changing from $H$ to $L$ increases the value of $e$ by $6$, decreases the value of each $b_j$ by $3$, and leaves unit values unchanged. Hence $L(S)\leq H(S)+6$ for every bundle $S$, giving
\[
 \E[L(X_1)]\leq\E[H(X_1)]+6\leq11.
\]

\paragraph{Step 2: bound a truthful $A$-agent's utility against two $H$-reports.}
Now consider the report profile $(A,H,H)$, and let $Y_i$ denote agent $i$'s random bundle, for $i=1,2,3$. Fairness forces agent $1$, who reports $A$, to receive $e$. Indeed, if an $H$-reporter received $e$, that agent would need all six units, leaving the $A$-reporter a $b_j$ without the unit she needs. Thus $Y_1$ contains $e$, neither $b_j$, and some unit goods.

To bound the value of $Y_1$, return to a \emph{true $L$-agent} facing two $H$-reports. Truthful reporting gives her $X_1$ from Step 1; reporting $A$ gives her $Y_1$. Her TIE constraint is
\[
 \E[L(Y_1)]\leq\E[L(X_1)]\leq11.
\]
On bundles of the form $Y_1$, valuations $A$ and $L$ differ only on $e$, where $L(e)-A(e)=4+\varepsilon$. Therefore
\[
 \E[A(Y_1)]
 =\E[L(Y_1)]-4-\varepsilon
 \leq11-4-\varepsilon=7-\varepsilon.
\]
Now suppose agent $1$'s true valuation is $A$ and she reports truthfully. The reports are again $(A,H,H)$, so the mechanism gives her the same lottery $Y_1$. Her truthful utility is now measured by $A$, and the calculation above bounds it by $7-\varepsilon$.

\paragraph{Step 3: some true $A$-agent gains by reporting $H$.}
Let $Z_1,Z_2,Z_3$ be the random bundles at $(H,H,H)$. The recipient of $e$ needs all six unit goods, and the other agents receive one $b_j$ each. Fairness therefore forces complete allocation at this profile. Evaluating all bundles using $A$ gives
\[
 \sum_{i=1}^3\E[A(Z_i)]=A(G)=21-\varepsilon.
\]
Thus some named agent $i$ has $\E[A(Z_i)]\geq7-\frac{\varepsilon}{3}$.

Now let this agent's true valuation be $A$, with both opponents reporting $H$. If she reports $H$, the profile is $(H,H,H)$, so her deviation utility is precisely $\E[A(Z_i)]$. If she reports truthfully, she receives a bundle $\widehat{Y}_i$ at the profile with her report $A$ and both other reports $H$. Steps 1 and 2 can be repeated with any named agent in the distinguished position; they do not rely on anonymity. They therefore bound this selected agent's truthful utility by $\E[A(\widehat{Y}_i)]\leq7-\varepsilon$. Hence
\[
 \underbrace{\E[A(\widehat{Y}_i)]}_{\text{truthful}}
 \leq7-\varepsilon
 <7-\frac{\varepsilon}{3}
 \leq\underbrace{\E[A(Z_i)]}_{\text{reporting }H},
\]
contradicting TIE. This proves the $\frac{6}{7}$ upper bound.

\paragraph{Extension to approximate TIE.}
We repeat the same comparisons with multiplicative slack $0\leq\delta<1$. The fairness-based bundle restrictions remain unchanged. In Step 1, the two true $H$-agents' constraints give
\[
 \E[H(X_2)],\E[H(X_3)]\geq8(1-\delta).
\]
Thus $\E[H(X_1)]\leq21-16(1-\delta)=5+16\delta$, and $\E[L(X_1)]\leq11+16\delta$. In Step 2, a true $L$-agent's deviation to $A$ gives
\[
 (1-\delta)\E[L(Y_1)]\leq\E[L(X_1)]\leq11+16\delta.
\]
Since $A(Y_1)=L(Y_1)-4-\varepsilon$, every named agent $i$ has truthful $A$-utility against $H,H$ bounded by
\[
 \E[A(\widehat{Y}_i)]\leq\frac{11+16\delta}{1-\delta}-4-\varepsilon.
\]
For the agent selected in Step 3, reporting $H$ still yields expected true value at least $7-\frac{\varepsilon}{3}$. Her approximate-TIE constraint therefore requires
\[
 (1-\delta)\left(7-\frac{\varepsilon}{3}\right)
 \leq\E[A(\widehat{Y}_i)]
 \leq\frac{11+16\delta}{1-\delta}-4-\varepsilon.
\]
Multiplying by $3(1-\delta)>0$ and collecting terms gives
\[
 2\varepsilon\leq(102+\varepsilon)\delta-(21-\varepsilon)\delta^2
 \leq(102+\varepsilon)\delta.
\]
Thus $\delta\geq\frac{2\varepsilon}{102+\varepsilon}$. The condition $\alpha(7-\varepsilon)>6$ permits every $0<\varepsilon<\eta$, where $\eta=7-\frac{6}{\alpha}\leq1$. Letting $\varepsilon$ approach $\eta$ from below yields
\[
 \delta\geq\frac{2\eta}{102+\eta},\qquad \eta=7-\frac{6}{\alpha}.
\]
Free disposal was allowed throughout; completeness at $(H,H,H)$ follows from fairness on that profile alone.\hfill$\square$

%% file: sections/general-goods.tex
\section{\texorpdfstring{$n\geq3$}{n >= 3}: an upper bound of \texorpdfstring{$\frac{9}{10}$}{9/10} MMS for TIE mechanisms}\label{sec:general-goods}

\begin{theorem}\label{thm:general-goods}
For every $n\geq3$, no TIE mechanism without payments guarantees more than $\frac{9}{10}$-MMS ex post for equally entitled, risk-neutral agents with nonnegative additive valuations on $n+5$ goods. This holds even with free disposal and a publicly known common strict ranking, with all reports required to respect that ranking.
\end{theorem}
We first prove the bound on a two-report domain and derive the quantitative incentive-slack bound. We then perturb the reports to establish the public-ranking strengthening.

We adapt the homogeneous-profile and unilateral-deviation argument of Lin, Tao, Wu, and Zhang~\cite[Theorem 5.3]{lin-et-al-chores} to goods. Their proof uses an eleven-core-chore construction and a bundle obstruction established in their Lemma 5.4. Here we use an eight-core-good construction and prove the corresponding bundle obstruction below.

\paragraph{Roadmap: one bundle lemma and one deviation.}
The incentive comparison uses the same true valuation $v$ throughout. The only change is one agent's report, from $v$ to $w$. The instance below defines these reports and the positive parameter $\varepsilon$.
\begin{center}
\renewcommand{\arraystretch}{1.3}
\begin{tabular}{p{0.43\linewidth}p{0.47\linewidth}}
\toprule
Report profile&Consequence, evaluated using $v$\\\midrule
Everyone reports $v$.&Total value is $10n-\varepsilon$, so some named agent has expected utility at most $10-\frac{\varepsilon}{n}$.\\
That agent reports $w$; all opponents still report $v$.&The bundle lemma below forces her value to be at least $10$ in every outcome.\\
TIE compares these two reports.&Would require $10-\frac{\varepsilon}{n}\geq10$: a contradiction.\\
\bottomrule
\end{tabular}
\end{center}
We begin with the instance and the fairness thresholds used in the lemma.

\paragraph{The instance.} Fix $n\geq3$ and $\alpha>\frac{9}{10}$ with $\alpha\leq1$. Suppose a TIE mechanism guarantees $\alpha$-MMS ex post. Choose $\varepsilon\in(0,1)$ so that $\alpha(10-\varepsilon)>9$. There are eight core goods and $n-3$ extra goods. The two valuations are:
\begin{center}
\begin{tabular}{lrrrrrr}
\toprule
&$a_1,a_2$&$b_1,b_2,b_3$&$d_1$&$d_2$&$e$&\shortstack{Each of the $n-3$\\extra goods}\\\midrule
$v$&$1$&$3$&$6-\varepsilon$&$6$&$7$&$10$\\
$w$&$1$&$3$&$4$&$5$&$7$&$9$\\\bottomrule
\end{tabular}
\end{center}
Entries are per item. In the eventual deviation, $v$ is the true valuation and $w$ is the alternative report. For the bundle calculations below, we temporarily ignore the $\varepsilon$ reduction in the value of $d_1$. Let $u(S)$ denote the resulting integer value of a bundle $S$. Thus $u(S)=v(S)+\varepsilon$ if $d_1\in S$, and $u(S)=v(S)$ otherwise.

\paragraph{Step 1: compute the MMS values and the resulting bundle requirements.}
The following partitions of the core, together with one singleton bundle for each extra good, give partitions into $n$ bundles:
\begin{center}
\begin{tabular}{l|ccc|c}
&\multicolumn{3}{c|}{Core bundles}&MMS lower bound\\\hline
$v$&$\{a_1,b_1,d_1\}$&$\{a_2,b_2,d_2\}$&$\{b_3,e\}$&$10-\varepsilon$\\
$w$&$\{a_1,a_2,e\}$&$\{b_1,b_2,b_3\}$&$\{d_1,d_2\}$&$9$
\end{tabular}
\end{center}
For $w$, all these bundles have value $9$, and $w(G)=9n$. The average-value upper bound therefore gives $\mu_w=9$.

For $v$, the displayed bundles have values $10-\varepsilon,10,10$ on the core and $10$ on each extra good, so $\mu_v\geq10-\varepsilon$. To prove equality, suppose a partition gave every bundle $v$-value strictly above $10-\varepsilon$. Since $0<\varepsilon<1$, every bundle would then have integer $u$-value at least $10$. But $u(G)=10n$, so every bundle would have $u$-value exactly $10$. The bundle containing $d_1$ would consequently have $v$-value exactly $10-\varepsilon$, a contradiction. Thus $\mu_v=10-\varepsilon$.

Now apply the assumed ex-post $\alpha$-MMS guarantee. A bundle $S$ assigned to a $v$-reporter satisfies $v(S)\geq\alpha(10-\varepsilon)>9$, hence $u(S)\geq10$ by integrality. A bundle $Y$ assigned to a $w$-reporter satisfies $w(Y)\geq9\alpha>8$, hence $w(Y)\geq9$. These requirements hold separately in every allocation in the mechanism's support.

\paragraph{Step 2: the bundle lemma.}
\begin{lemma}\label{lem:forced-bundle}
At a profile with one report $w$ and $n-1$ reports $v$, every allocation satisfying the ex-post $\alpha$-MMS guarantee gives the $w$-reporter a bundle $Y$ with $v(Y)\geq10$. This holds for every choice of the reporting agent and allows free disposal.
\end{lemma}
\begin{proof}
Fix an allocation satisfying the fairness guarantee, and let $Y$ be the $w$-reporter's bundle. Fairness gives $w(Y)\geq9$. Our goal is to show that this same bundle has value at least $10$ under $v$, the true valuation used in the deviation argument.

If $Y$ contains an extra good, then $v(Y)\geq10$, as required, since each extra good has $v$-value $10$. It remains to consider the case where $Y$ contains only core goods. Suppose, towards a contradiction, that $v(Y)<10$.

\emph{(i) The bundle must have $u(Y)=w(Y)=9$.}
Since $u(Y)\leq v(Y)+\varepsilon<11$ and $u(Y)$ is an integer, $u(Y)\leq10$. We first rule out equality. If $u(Y)=10$, then $Y$ must contain $d_1$: without $d_1$, we would have $v(Y)=u(Y)=10$. But $d_1$ is worth $6$ under $u$ and only $4$ under $w$, and every other good is worth no more under $w$ than under $u$. Thus $w(Y)\leq u(Y)-2=8$, contradicting fairness. Hence $u(Y)\leq9$. Together with fairness and $w\leq u$, this gives
\[
 9\leq w(Y)\leq u(Y)\leq9,
 \qquad\text{so}\qquad w(Y)=u(Y)=9.
\]

\emph{(ii) There are only two possible bundles with these values.}
Including $d_1$ would make $u(Y)-w(Y)$ at least $2$, and including $d_2$ would make it at least $1$. Neither difference can be offset by another good, since $u\geq w$ item by item. The equality $u(Y)=w(Y)$ therefore excludes both $d_1$ and $d_2$.

The remaining candidates are $a_1,a_2,b_1,b_2,b_3,e$, with values $1,1,3,3,3,7$ under both $u$ and $w$. Their selected values must sum to exactly $9$:
\begin{itemize}
\item If $e\in Y$, its value is $7$, leaving exactly $2$ to be supplied. No $b_j$ can be included, since it is worth $3$. The only possibility is to include both $a_1$ and $a_2$. Thus $Y=\{a_1,a_2,e\}$.
\item If $e\notin Y$, using at most two $b$-goods gives value at most $3+3+1+1=8$. Thus all three $b$-goods must be included. They already total $9$, so neither $a$-good can be included. Thus $Y=\{b_1,b_2,b_3\}$.
\end{itemize}

\emph{(iii) The remaining core must serve two agents, each needing value at least $10$.}
There are $n-1$ other agents and only $n-3$ extra goods. At least two of these agents therefore receive no extra good; their bundles, say $R$ and $T$, come entirely from the core outside $Y$. Both agents report $v$, so Step 1 gives $u(R),u(T)\geq10$.

The whole core has $u$-value $30$, and $u(Y)=9$, leaving only $21$. Since $R$ and $T$ are disjoint,
\[
 u(R)+u(T)\leq21,
 \qquad 10\leq u(R),u(T)\leq11.
\]
At most $1$ unit of the remaining core value can lie outside $R\cup T$. Consequently, every remaining core good worth more than $1$ under $u$ must belong to $R$ or $T$. This conclusion allows both disposal and allocations of core goods to other agents.

\emph{(iv) Neither candidate leaves suitable bundles for these two agents.}
The two cases are summarized below; all remaining values are measured using $u$.
\begin{center}
\begin{tabular}{p{0.20\linewidth}p{0.24\linewidth}p{0.46\linewidth}}
\toprule
Possible $Y$&Remaining core $u$-values&Obstruction\\\midrule
$\{a_1,a_2,e\}$&$6,6,3,3,3$&Each of $R,T$ needs a $6$ and two $3$'s, but only three $3$'s remain.\\
$\{b_1,b_2,b_3\}$&$7,6,6,1,1$&The three large goods must go into $R,T$, forcing one bundle to have value at least $12$.\\\bottomrule
\end{tabular}
\end{center}

If $Y=\{a_1,a_2,e\}$, both value-$6$ goods must belong to $R\cup T$. They cannot go to the same bundle, since that bundle would have value at least $12$, exceeding the upper bound $11$. Thus each bundle receives exactly one $6$. One additional $3$ gives only $9$, so each bundle needs at least two $3$'s to reach $10$. That requires four value-$3$ goods, but only three remain.

If $Y=\{b_1,b_2,b_3\}$, all three goods of values $7,6,6$ must belong to $R\cup T$. Placing three goods into two bundles forces one bundle to contain at least two of them. Even the two smallest total $12$, again exceeding the upper bound $11$.

Both candidates are impossible, so $v(Y)\geq10$, proving the lemma.
\end{proof}

\paragraph{Step 3: compare truthful reporting with the deviation to $w$.}
Consider the true valuation profile $(v,\ldots,v)$, with every agent reporting truthfully, and let $X_i$ be agent $i$'s random bundle. Disjointness and nonnegative values imply
\[
 \sum_{i=1}^n\E[v(X_i)]\leq v(G)=10n-\varepsilon.
\]
Thus some named agent $i$ has truthful expected utility at most
\[
 \E[v(X_i)]\leq10-\frac{\varepsilon}{n}.
\]
Keep this agent's \emph{true valuation equal to $v$} and the other reports fixed at $v$, but let her report $w$. Write $Y_i$ for her resulting random bundle. Lemma~\ref{lem:forced-bundle} applies to this selected agent and gives $v(Y_i)\geq10$ in every outcome, so $\E[v(Y_i)]\geq10$.

Exact TIE requires truthful expected utility to be at least deviation utility. Both are evaluated using $v$, so
\[
 10-\frac{\varepsilon}{n}\geq\E[v(X_i)]
 \geq\E[v(Y_i)]\geq10,
\]
which is impossible for $\varepsilon>0$. No anonymity or complete-allocation assumption was used: averaging selects the agent, and the bundle lemma applies to every agent.

\paragraph{Extension to approximate TIE.}
The bundle lemma uses only fairness, so it remains valid with incentive slack $0\leq\delta<1$. For the same selected true $v$-agent facing $v$-reports, approximate TIE requires
\[
 10-\frac{\varepsilon}{n}\geq\E[v(X_i)]
 \geq(1-\delta)\E[v(Y_i)]\geq10(1-\delta).
\]
Rearranging gives $\delta\geq\frac{\varepsilon}{10n}$.
The condition $\alpha(10-\varepsilon)>9$ allows every $0<\varepsilon<10-\frac{9}{\alpha}$, and $10-\frac{9}{\alpha}\leq1$ because $\alpha\leq1$. Letting $\varepsilon$ approach $10-\frac{9}{\alpha}$ from below yields
\[
 \delta\geq\frac{10-\frac{9}{\alpha}}{10n}
 =\frac{10\alpha-9}{10\alpha n}.
\]
\paragraph{Separation even where exact MMS always exists.}
Feige, Sapir, and Tauber~\cite{feige-sapir-tauber} show that exact MMS exists at every additive profile on at most $n+5$ goods. Our witness uses exactly this many goods, so its loss is due to incentives alone. Together with the four-good, two-agent construction, this gives a separation on an exact-MMS item domain for every $n\geq2$. The comparison for unrestricted item sets is summarized in Section~\ref{sec:model}.

\paragraph{Strengthening to a publicly known common strict ranking.}
We now establish the public-ranking claim in Theorem~\ref{thm:general-goods}. All reports must respect the same publicly known strict ranking, while cardinal values remain private~\cite[Section 5]{amanatidis-2016}.

The reports above have ties, which we remove by a small perturbation. If arbitrarily small perturbations admitted mechanisms with the required guarantees, their limiting lotteries would contradict the unperturbed proof. This argument does not assume continuity of the mechanism.

\emph{(i) Give both reports the same strict ranking.} Fix $n\geq3$ and $\frac{9}{10}<\alpha\leq1$, and choose $\varepsilon$ as above. The reports $v,w$ have a compatible weak ranking: the extra goods come first, followed by $e,d_2,d_1$, the three $b$-goods, and the two $a$-goods. Fix an order within each tied group and write the resulting item order as $g_1,\ldots,g_m$, where $m=n+5$. For $\tau>0$, define
\[
 v^\tau(g_j)=v(g_j)+(m+1-j)\tau,
 \qquad w^\tau(g_j)=w(g_j)+(m+1-j)\tau.
\]
An earlier good receives a larger increment. Thus a strict comparison remains strict, and every tie is broken in the chosen direction. Both reports have the same strict ranking $g_1\succ\cdots\succ g_m$, independently of $\tau$. Reporting $w^\tau$ instead of $v^\tau$ is therefore an admissible deviation even when this ranking is public. As $\tau\to0$, each item value, and hence each bundle value, converges to its unperturbed value.

\emph{(ii) Suppose arbitrarily small perturbations admit a mechanism.} A TIE, ex-post $\alpha$-MMS mechanism on the full domain respecting the public ranking would retain both properties when restricted to $\{v^\tau,w^\tau\}^n$: all profiles and deviations in this smaller domain are still admissible. It is therefore enough to rule out mechanisms on these finite domains for sufficiently small $\tau$.

Suppose, towards a contradiction, that there is a sequence $\tau_k\to0$ of positive perturbations, each admitting such a mechanism. The mechanism may be different for each $k$. We keep $n$, the item set, $\alpha$, and $\varepsilon$ fixed throughout.

\emph{(iii) Choose a subsequence on which every allocation probability converges.} Let $\mathcal D$ be the finite set of allocations, including those with discarded goods. For an unperturbed report profile $Q\in\{v,w\}^n$, let $Q^{\tau_k}$ denote its perturbed version, and write
\[
 p_k(D\mid Q)=\Pr[\text{the $k$-th mechanism outputs }D
                  \text{ at }Q^{\tau_k}],\qquad D\in\mathcal D.
\]
There are only $2^n$ profiles and finitely many allocations, so these probabilities form a vector with finitely many coordinates, all in $[0,1]$. A subsequence therefore converges in every coordinate simultaneously. Along this subsequence, write its limits as $p(D\mid Q)$. For each $Q$, nonnegativity and the finite sum give
\[
 p(D\mid Q)\geq0,\qquad
 \sum_{D\in\mathcal D}p(D\mid Q)
 =\lim_{k\to\infty}\sum_{D\in\mathcal D}p_k(D\mid Q)=1.
\]
These probabilities define a valid lottery at every original profile $Q$.

\emph{(iv) Every incentive constraint holds in the limit.} Fix agent $i$, a true valuation $T\in\{v,w\}$, an alternative report $S\in\{v,w\}$, and opponents' reports $Q_{-i}$. Write $D_i$ for agent $i$'s bundle in allocation $D$. TIE for the $k$-th mechanism gives
\[
 \sum_{D\in\mathcal D}p_k(D\mid T,Q_{-i})\,T^{\tau_k}(D_i)
 \geq
 \sum_{D\in\mathcal D}p_k(D\mid S,Q_{-i})\,T^{\tau_k}(D_i).
\]
On the left, the agent reports her true valuation $T^{\tau_k}$; on the right, she reports $S^{\tau_k}$. Both utilities are measured using $T^{\tau_k}$, and the opponents' perturbed reports are fixed.

Each allocation probability converges by Step (iii), and each true bundle value converges to $T(D_i)$ by Step (i). Since the sums are finite, both expected utilities converge. An inequality between two convergent sequences also holds between their limits, so
\[
 \sum_{D\in\mathcal D}p(D\mid T,Q_{-i})\,T(D_i)
 \geq
 \sum_{D\in\mathcal D}p(D\mid S,Q_{-i})\,T(D_i).
\]
This is the TIE constraint for the limiting lotteries. Since $i,T,S,Q_{-i}$ were arbitrary, all incentive constraints on $\{v,w\}^n$ hold, including the deviation from $v$ to $w$ used above.

\emph{(v) Every allocation with positive limiting probability is fair.} First, MMS converges with the item values. For each fixed partition, all bundle values converge, so their minimum converges. There are only finitely many partitions; taking their maximum therefore also preserves convergence. In particular,
\[
 \mu_{T^{\tau_k}}\longrightarrow\mu_T
 \qquad\text{for }T\in\{v,w\}.
\]
Fix a profile $Q$ and an allocation $D$ that is \emph{not} $\alpha$-MMS for $Q$. Some agent $i$ has a strictly positive shortfall
\[
 r=\alpha\mu_{Q_i}-Q_i(D_i)>0.
\]
The perturbed shortfall converges to the same positive number:
\[
 \alpha\mu_{Q_i^{\tau_k}}-Q_i^{\tau_k}(D_i)\longrightarrow r.
\]
For all sufficiently large $k$, it is therefore greater than $\frac{r}{2}>0$. Hence $D$ also violates fairness at $Q^{\tau_k}$. Ex-post fairness of the $k$-th mechanism forces $p_k(D\mid Q)=0$ for all these $k$, and consequently $p(D\mid Q)=0$.

Every allocation with positive limiting probability is therefore $\alpha$-MMS, establishing the required ex-post guarantee.

\emph{(vi) Apply the unperturbed contradiction.} The limiting lotteries define a TIE, ex-post $\alpha$-MMS mechanism on $\{v,w\}^n$. But the bundle lemma and the incentive comparison proved above use only profiles and deviations in this domain. They force some true $v$-agent to have truthful expected utility at most $10-\frac{\varepsilon}{n}$ and deviation utility at least $10$, contradicting TIE.

Therefore arbitrarily small perturbations cannot admit such mechanisms: for all sufficiently small positive $\tau$, the impossibility holds on the perturbed domain. These reports share a strict public ranking, proving the stronger claim. The perturbation uses the same $n+5$ goods, so exact MMS still exists at every profile in this restricted domain.

For more than $n+5$ goods, first append goods valued at zero under both original reports and place them last in the common weak order. These goods do not change the unperturbed contradiction. Applying the same perturbation with the enlarged value of $m$ and repeating Steps (ii)--(vi) proves the public-ranking impossibility on the enlarged item set as well.\hfill$\square$

%% file: sections/ai-disclosure.tex
\section*{AI disclosure}
The results presented here were obtained by GPT Astra 6.2 by adapting the constructions of Lin et al.~\cite{lin-et-al-chores} to goods. GPT Astra 6.2 also drafted portions of the article. The authors have verified the proofs and take full responsibility for the correctness of the results and for the content of the article.